# Can we improve the energy efficiency of EUV lithography?

Tsumoru Shintake

OIST: Okinawa Institute of Science & Technology Graduate University,
1919-1 Tancha, Onna-son, Okinawa, Japan 904-0495

## ABSTRACT

This paper discusses a simple, low-cost, highly efficient two-mirror projector with a simplified illumination system. The EUV source power can be reduced by 1/10 compared to the current six-mirror EUV projector system. The required EUV power is 20 watts for process speed of 100 wafers per hour. The proposed in-line projector achieves 0.2 NA (20 mm field) and 0.3 NA (10 mm field), which can be assembled in a cylindrical tube configuration similar to a DUV projector, providing superior mechanical stability and easier assembly/maintenance. The EUV light is introduced in front of the mask through two narrow cylindrical mirrors located on both side of the diffraction cone, providing average normal illumination and reducing the mask 3D effect. The simplified illumination system provides symmetric quadrupole off-axis illumination, bypassing central obscuration and improving spatial resolution, also realizing Köhler illumination. The theoretical resolution limit is 24 nm (20 mm field), image reduction factor x5 and object image distance (OID) 2000 mm. With the curved surface mask, the tool height can be reduced to (OID) 1500 mm, which provides resolution 16 nm (10 mm field). It will be suitable for small die size chip production for mobile applications as well as the latest chiplet technology.

## 1. INTRODUCTION

Extensive research and development, as well as significant investment, have been dedicated to EUV lithography over the past few decades. Key components, such as the high-precision multilayer mirror with a diameter of nearly 1 meter and the high-power EUV light source, have been successfully developed [1]. Production lithography tools with a numerical aperture (NA) of 0.33 are currently in operation. However, for EUV lithography to be widely accepted as a reliable tool for high-volume fabrication, it must be economically feasible. Therefore, the issue of cost needs to be addressed. While Moore's law continues to hold true, it is important to remember that the resources available on our planet are limited. Therefore, we must strive to meet the sustainable development goals. The chip industry should avoid excessive consumption of electricity and water in the development of the next generation.

This paper aims to find a cost-effective solution that meets performance requirements using available technologies within a reasonable timeframe. Therefore, we concentrate on the low-NA lithography of inline two-mirror configuration, as illustrated in Fig. 1. This approach would help to reduce costs and save electricity consumption.

The multi-layer mirror absorbs more than 30% of EUV power at each reflection [2]. Current exposure tools have six mirrors in the projection optics and four mirrors in the illuminator, the power transmission from the EUV source to the wafer is quite low. In contrast, in the two-mirror projector with simplified illuminator proposed in this paper using two series mirrors, the power transfer efficiency will be drastically improved.

$$\text{This proposal: } P_t(2+2) = 0.65^4 = 0.18 \tag{1a}$$

$$\text{Current EUV tool: } P_t(4+6) = 0.65^{10} = 0.013 \tag{1b}$$

The proposal is 13 times more efficient, resulting in a 92% reduction in electricity consumption for EUV power generation. This will lower the AC power consumption from approximately 1 MW to 80 kW. Additionally, the cooling water flow in the drive laser system will be significantly reduced. The required EUV power at intermediate focus is 20 W for 100 wafer-per-hour throughput per tool. The design of the EUV source is simplified, resulting in reduced investment and maintenance costs and improved reliability. At this power level, a thin film window, similar to the pellicle on the mask, can be placed in the illumination system around the intermediate focus (IF) to prevent debris from the plasma source, thus protecting the expensive mask and mirrors. Existing EUV tools have typically had slower scanning speeds than optical scanners due to the weakness of EUV light sources. However, by using the system proposed in this paper, we can increase the EUV power to the wafer, resulting in faster scanning speeds and increased productivity.

The surface roughness of the projector mirrors affects the image quality. For this reason, extensive technical R&D has been undertaken to achieve an ultra-precise surface. In particular, mid-frequency roughness in micrometer range severely affects image contrast, the same phenomenon as fog in nature. In synchrotron light facilities, carbon contamination often occurs on the surface of mirrors, which also causes loss of image contrast. The surface cleanliness inside the EUV lithography tank can be much better, while we should be careful about carbon contamination. It is advisable to reduce the number of mirrors in the projector to achieve a high contrast image and maintain it for a long time.

It should be also noted that the heating of the wafer due to the infrared light transmitted through the optics from the CO2 drive laser of the EUV source affects the overlay control. In the proposed system, the EUV source power and the CO2 drive laser power have been reduced by a factor of ten, thus eliminating this problem.

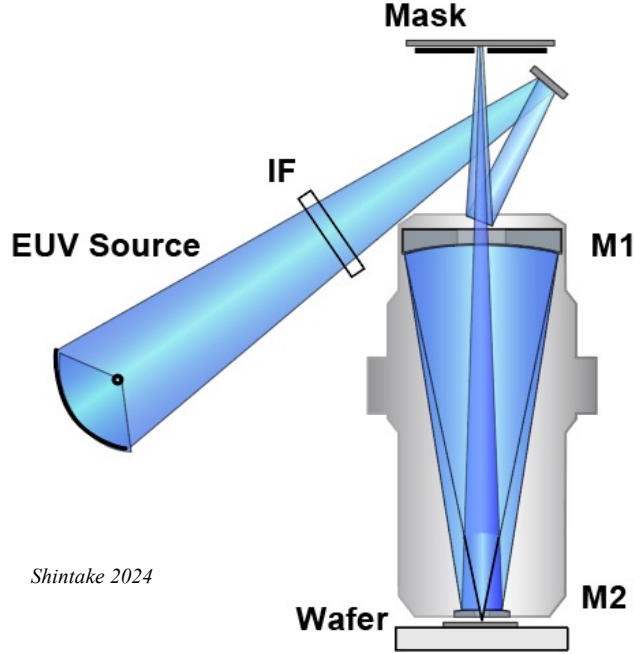

Figure 1. In-line two-mirror projector with simplified illuminator. The number of mirrors is much lower than in the current EUV lithography system, so the power transmission can be greatly improved.

At lower NA, optical aberration correction is easier because the light rays run near the axis. Only two aspheric mirrors are needed to cover a reasonably wide image field. The optics simulation has confirmed that NA 0.2 will provide an image field of 20 mm in size for a 2 m high projector. Compared to immersion iArF, low-NA EUV offers superior resolution due to its shorter wavelength of 13.5 nm, which is 15 times shorter than ArF's 198 nm. The critical dimension or resolution is determined by the Abbe equation:

$$CD = k_1 \cdot \frac{\lambda}{NA} \qquad (2)$$

where $k_1$ represents the process factor, $\lambda$ represents the wavelength, and NA represents the numerical aperture. The spatial resolution is determined in two cases:

$$CD_{\text{EUV 0.2NA}} = 24 \text{ nm} \qquad (3a)$$

$$CD_{\text{iArF 1.35NA}} = 40 \text{ nm} \qquad (3b)$$

where $k_1$ is equal to 0.36 and 0.27 is assumed in EUV and iArF cases, respectively. It may be possible to achieve a single patterning on a 24 nm half-pitch using this low-NA EUV. Refer to the later section on $k_1 = 0.35$ in EUV.

The other important issue is the depth of focus (DOF), which is defined as follows.

$$DOF = k_2 \cdot \frac{\lambda}{(NA)^2}$$

(4)

Using eq. (2) into eq. 43), we find the dimensionless relation:

$$\frac{DOF}{CD} \approx \frac{1}{NA}$$

(5)

This equation tells us lower-NA always provides longer DOF. At two cases,

$$DOF_{\text{EUV 0.2NA}} = 340 \text{ nm}$$

(6a)

$$DOF_{\text{iArF 1.35NA}} = 110 \text{ nm}$$

(6b)

In both cases, we assume $k_2 = 1$. It is clear that the low-NA EUV has an advantage for longer DOF. Furthermore, compared to common EUV projectors that use oblique illumination on the mask, the inline projector does not exhibit EUV-specific image variations around the focus due to average-normal illumination. This eliminates image-placement errors caused by mask non-flatness. The use of low-NA EUV therefore simplifies the mask and wafer flatness and focus control requirements. It also makes it easier to realize the curved surface mask, which will be discussed in a later section.

Axisymmetric optics provide uniform image contrast around the axis, simplifying source mask optimization (SMO). Conventional quadrupole illumination will be sufficient. In addition, the maximum reflection angle on AM2 is only 5.5 degrees from the surface normal. This results in minimal asymmetric pupil apodization, no polarization dependence, and no phase changes associated with multi-layer coating.

At the EUV wavelength, it is important to consider the quantum mechanical effect, specifically the potential for higher photon energy to degrade patterning, which is known as the stochastic effect. The photon energy is given by

$$E = h \cdot \nu = hc/\lambda = 1240/\lambda \text{ (eV/nm)}$$

(7)

At both cases:

$$E_{\text{EUV 13.5 nm}} = 92 \text{ eV}$$

(8a)

$$E_{\text{ArF 198 nm}} = 6.5 \text{ eV}$$

(8b)

EUV photons have 14 times more energy than ArF, resulting in 14 times fewer photo-ionization events within the resist layer at the same absorption energy. This leads to worse LER (line edge roughness), due to random Poisson distribution. The lithographic capability is limited by defects caused by stochastic phenomena. We have to remined that a contact failure rate of less than 3 x $10^{-11}$ is necessary for the large-scale production of a logic circuit[3]. A number of research and development teams are currently working on the understanding of the mechanisms involved, and new types of resistant materials have been proposed to overcome these challenges. However, it may take some time before these solutions become available. In the meantime, it is advisable to use a projector with a lower NA and working on wider line spacing. Lower cost EUV lithography may push multiple patterning to narrower line widths [4]. It is also important to note that a two-mirror projector can provide more photons and help reduce statistical noise.

As shown in Fig. 1, the two-mirror projector is mounted in a tube similar to those used in ultraviolet lithographic lenses. Extremely high-precision mirrors are encapsulated in the tube, forming a monolith that offers several advantages, including mechanical stability, ease of assembly, alignment, replacement and superior sealing to protect against dust contamination. This results in lower capital and maintenance costs and improved reliability.

## 2. ABERRATION CORRECTED OPTICS

### 2.1 Aberration correction in two-mirror equal radii configuration

EUV photolithography requires a flat field anastigmat that uses only reflective mirrors. The Petzval sum rule is the central principle for the flat field projector. In a two-mirror configuration,

$$\sum_{k=1}^{n} \frac{2}{R_i} = \frac{2}{R_1} - \frac{2}{R_2} = 0 \tag{9}$$

where $R$ is the mirror curvature. The most basic configuration for two-mirror projector should consist of positive and negative power mirrors, specifically a concave and a convex mirror with the same radius. This is known as the 'equal radii' configuration, as illustrated in Fig. 2. When the distance between two mirrors is L = 0.86 R, the object (OBJ) on the secondary mirror M2 is projected onto the image (IMG) on the first mirror M1, correcting 3rd-order spherical aberration [5].

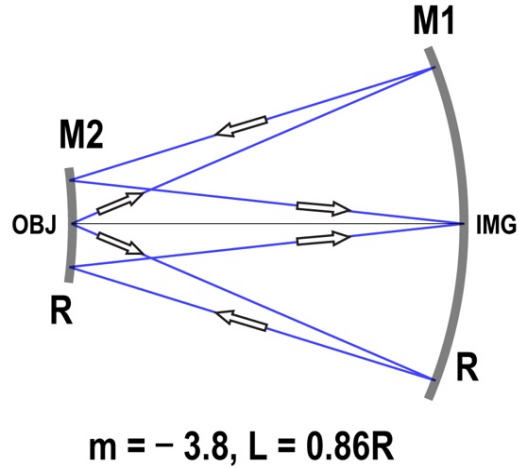

Figure 2. The equal radii configuration.

To create a functional projector, we adjust the mirror curvatures to pull the vertices (OBJ and IMG) outwards through the central holes. This will cause a break in the Petzval-sum rule, resulting in aberration. Aspheric mirrors have to be introduced to correct aberrations, but the numerical aperture and field size are limited due to the limited number of free aspheric parameters (only two mirrors available).

The equal radii configuration, named MET: Micro-Exposure Tool was carefully studied as the EUV projector by R. M. Hudyma and R. Soufli in 2008 [6]. The MET with a numerical aperture (NA) of 0.3 was designed to demonstrate a 30-nanometer half-pitch imaging. One of their designs assumed a virtual transmission mask and an inline projector, configured similarly to Fig. 1, but the illumination has to be provided through the virtual transmission mask. Aspheric mirrors were used to correct the aberrations, resulting in a residual root-mean-square (rms) wavefront error of 0.027λ. The projector was compact and had an object image distance (OID) of 276 mm. However, it was unsuitable as a lithography tool because its field of view was limited to 0.6 mm x 0.2 mm.

In 2004, using synchrotron radiation at the Advance Light Source in Berkeley, 30nm equal-line-space printing was demonstrated at MET [7]. This success suggested great potential for the two-mirror projector.

### 2.2 Extending field size.

To increase the field size, the length of the projector needs to be increased. It is assumed that the tool height is within the maximum size acceptable in a real semiconductor foundry:

$$OID = 2000 \text{ mm} \tag{10}$$

To maintain the Petzval-sum rule, it is critical to position the mirror M2 sufficiently close to the wafer. By assuming the same gap size between lens-wafer as the ArF immersion, it is recommended that the gap between the wafer and M2 mirror body should be 5 mm. To ensure the mirror body remains rigid, the distance between the wafer and M2 surface should be longer than 40-50mm. As demonstrated below, two mirrors with very similar curvatures (within 0.3% of each other) result in a wider field of view.

$$R(M1) = 1500 \text{ mm} \tag{11a}$$

$$R(M2) = 1505 \text{ mm} \tag{11b}$$

OpTaLix simulator [7] predicted a field of 20 mm at NA = 0.2, which covers the full mask-field of 100 mm. The image reduction factor is 1/5. We may also introduce a curved surface mask to cure residual field curve for short tool height and reducing wavefront error as discussed later.

## 2.3 Practical design as the two-mirror projector

Optical ray simulation results are shown in Fig. 3, where AM1 and AM2 are axial-symmetric aspheric mirrors. To direct the illumination into the projector, a spacious room is required to accommodate the cylindrical mirrors between mirror AM1 and the mask. This results in a magnification factor of x5, which is equivalent to MET, instead of the standard magnification factor of x4. The mask scan field measures 100 mm (20 mm x 5), which matches the current mask design of 104 mm (26 mm x 4). The simulation results at NA 0.2 are summarized in Table-1 and Table-2.

The simulation assumes a perfect mirror surface with 100% reflectivity. In real, the mirror is made of multi-layer coating, where the reflection is caused by the wave interference between these layers, resulting in amplitude and phase shift as the reflection angle changes. We need further careful simulation, including multi-layer coating, which will result in a change in the aspheric curvature, although it will be quite small. In practice, we need to measure the mirror quality with the interferometer at visible wavelength.

The wafer side is telecentric, but the mask side is not. Therefore, the chief ray is tilted; 1.6 degrees at the edge of the field (~50 mm/2000 mm radian). Considering the half angle of the diffraction cone (NA/5= 0.04 radians = 2.4 degrees), the maximum angle of reflection from the multilayer coating at the mask edge becomes 4 degrees. The angle is smaller than the 12-degree cut-off angle of the Mo/Si multilayer coating, resulting in minimal contrast loss. The defocus still causes pattern shift, with a wafer height error of 100 nm leading to 3 nm shift at the field edge. This shift is considered acceptable.

Note that all light rays from different fields intersect at the focal plane, creating diffraction spots that represent Fourier space. The light must pass through the central hole on both mirrors, which obscures the central part of the diffraction signal. The effect of the central obscuration is estimated separately at the focal plane using Fourier analysis (see later section).

Figures 4 and 5 shows that the wavefront aberration and the spot diagram. The optical path difference has small errors in small image height, but at the edge of the field, it reaches the limit of 0.05 wavelength due to residual aberration. Since NA is low, Strehl ratio is still high (0.991) at the field edge. We have to note that the Strehl ratio is estimated without the central obscuration and on-axis illumination. If we tilt the illumination, the higher frequency components start to pass the projector and resolution becomes higher, as a result it becomes aberration dominant. Fortunately, the optical path difference in Fig. 4 is axi-symmetric (it is actually cylindrically symmetric), the phase difference between the 1st order Bragg diffraction (see Fig. 12) from the narrowest pattern and the off-axis quadruple illumination becomes smaller, which means it effectively reduces the aberration. Further detailed investigations are needed.

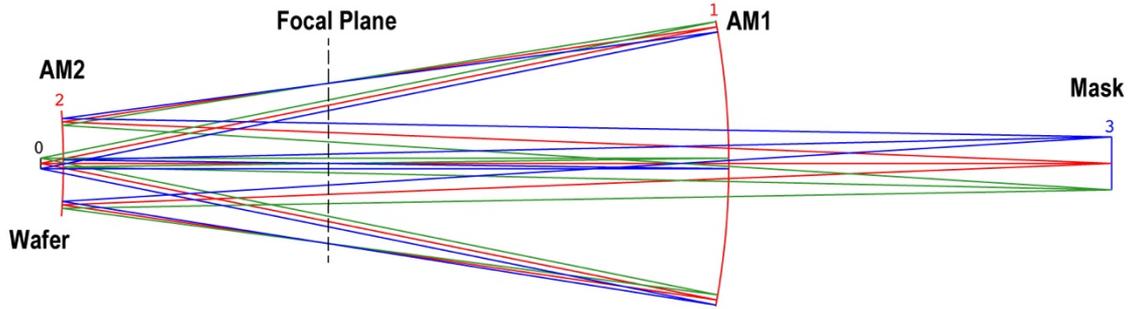

Figure 3. The simulation results for the inline two-mirror projector with 0.2 NA and OID of 2000 mm are presented. The mirrors are assumed to have a perfect surface with 100% reflectivity, and there is no pupil apodization or aperture stop. Note that in actual lithography, light travels in the opposite direction. To establish telecentric conditions on the OpTaLix simulator, it is easier to start the optical ray from the wafer side.

Table 1. Parameter list for the two-mirror projector.

| Parameter | Value | unit |
|---|---|---|
| NA | 0.2 | |
| Field size on the wafer | 20 | mm |
| RMS wavefront aberration at field edge (y=10 mm) | 0.015 | wave |
| Distortion at field edge (y=10 mm) | -0.026 | % |
| Strehl ratio (y=10 mm) | 0.991 | |
| Obscuration | $\Sigma_x = 0.13$ $\Sigma_y = 0.26$ | |
| AM1 curvature, diameter | 1500, 540 | mm |
| AM2 curvature, diameter | 1505, 180 | mm |
| Wafer to AM2 surface | 43 | mm |
| Telecentric | wafer side | |
| Magnification | x 5.0 | |
| Object to image distance (OID) | 2000 | mm |

Table-2: Aspheric prescription (OpTaLix output)

| Mirror | K | A | B | C | D |
|---|---|---|---|---|---|
| AM1 | 0 | -0.341E-11 | -0.175E-17 | -0.640E-23 | 0.178E-27 |
| AM2 | 0 | -0.402E-09 | -0.751E-15 | -0.444E-19 | 0.130E-22 |

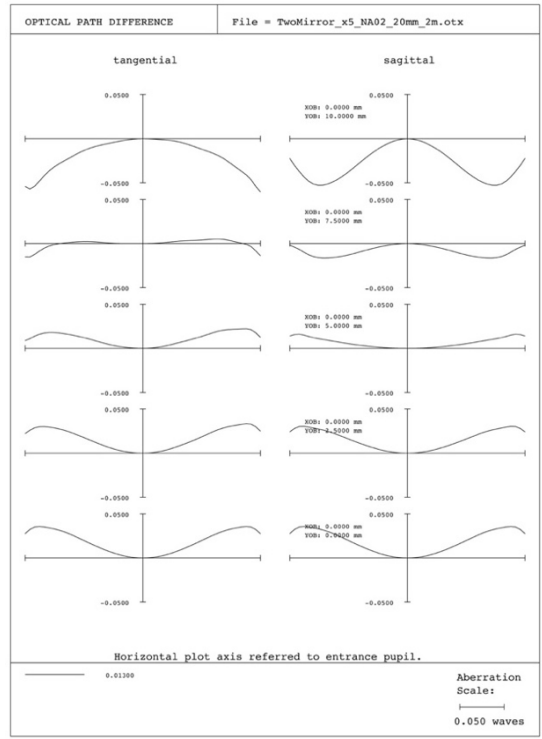

Figure 4. The graph shows the optical path difference along the beam height, with a vertical scale of 0.05 EUV wavelength 13.5 nm. At the scan field edge (y = 10 mm), Strehl ratio is as high as 0.991, resulting in a diffraction-limited spot at NA 0.2.

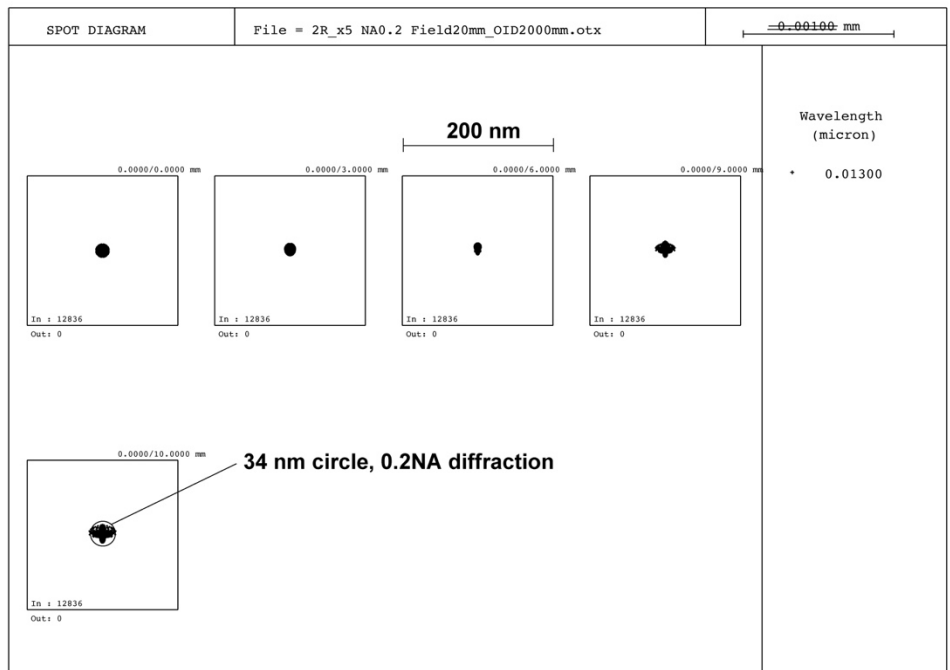

Figure 5. Spot diagram on the wafer. The output of the simulation is the spot on the mask, from which the spot on the wafer side can be estimated taking into account the image reduction factor of 1/5.

**2.4 Curved Surface Mask Option**

Due to the limited number of mirrors, the projected image is not perfectly flat and curved. The best focus point varies with field height, causing wavefront errors as shown in Fig. 4. If we introduce a curved surface mask as shown in Fig. 6, we can compensate y-field curve. We design the mask curvature to fit Petzval field curvature as follows,

$$\sum_{k=1}^{n} \frac{1}{R_i} = \frac{1}{R_1} + \frac{1}{R_2} + , , = \frac{1}{R_{curve}}$$

(12)

where $R_{curve}$ is the theoretical optimum radius of the curved surface mask. In practice, OpTaLix predicts a slightly smaller radius compensating spherical aberration term together.

By introducing the curved surface mask, which increases the freedom of design parameters, i.e. we can reduce the tool height and also increase the thickness of the AM2 mirror. Table-3 summarizes the design parameters with the curved surface mask that satisfies the Strehl ratio higher than 0.99. The bend radius is large and the amount of bend is relatively small compared to the mask width, there is no mechanical problem on the mask, and the transverse pattern shift can be integrated in the pattern design. We fabricate a flat mask as usual, then apply bend when the mask is mounted on the chucking on the scanner which has designed curve. Further discussion is required with the mask developer and pattern designer.

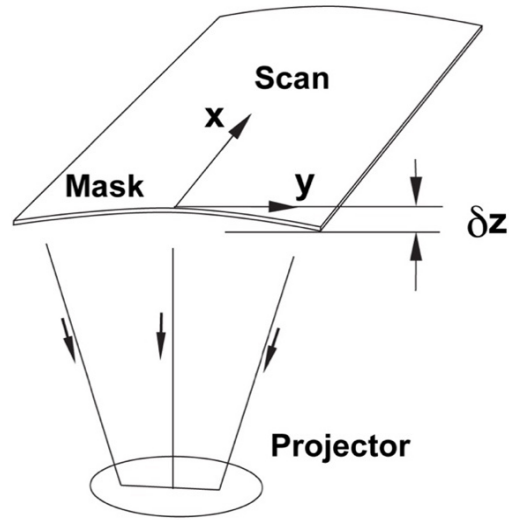

Figure 6. Curved surface mask concept compensating y-field curve.

Table-3: Design parameter with the curved surface mask.

| NA | Tool height | Curvature | Maximum bend at mask edge | Scan Field |
|---|---|---|---|---|
| 0.3 | 1.5 m | ~ 7m | -31 micron | 10 mm |
| 0.2 | 1.5 m | ~ 7m | -120 micron | 20 mm |

**2.5 Distortion**

As well known, the distortion causes image smearing in the scanner [9]. In two-mirror projector, the magnification decreases with axial distance causing the characteristic "barrel" distortion, which can be described mathematically:

$$r' = r[1 - C_d(r/r_0)^2]$$

(13a)

$$dr'/dr = 1 - 3C_d(r/r_0)^2 \approx 1 - 3C_d \qquad (13b)$$

The distortion $C_d$ is given in %. $r$ is the ideal paraxial position and $r'$ is the distorted position. As shown in Fig. 7, the linear scan motion of a point on the mask is projected as a curved trajectory (dashed line) due to the radial distortion. As discussed later, we use a dual line field, separated from the center. Since image smearing is less for the smaller separation, we minimize the separation as two scan widths that are tangent to each other on the axis.

The distortion effect shifts A point to A' point, and B point on edge to B' point as shown in Fig. 7. From height difference between A' and B', we find the smearing shift on the wafer. Using equations (13a), (13b) and

$$r \approx y(1 + x^2/2y^2) \qquad (14a)$$

$$\delta_y \approx \frac{3}{2m} C_d \ y_0 \cdot \frac{w^2}{y_0^2} \qquad (14b)$$

where $m$ is the image magnification factor $m = 5$. When we use $w = 2.5$ mm, $y_0 = 50$ mm on the mask, the smearing width becomes 9 nm on the wafer. The curved trajectory is more parallel to the scan motion in the central part and the rms value becomes roughly one-third: 3 nm. This will be acceptable level to print 24 nm half-pitch lines.

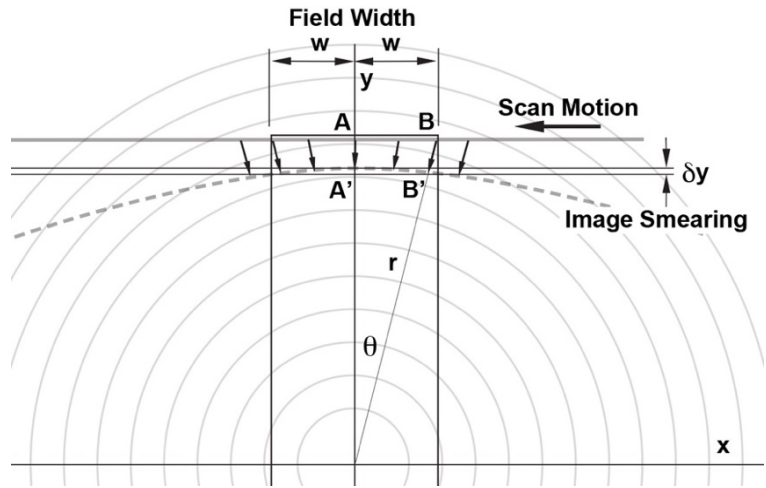

Figure 7. Image smearing due to the radial distortion during scan motion.

## 3. CENTRAL OBSCURATION

The two-mirror inline projector design inevitably has the problem of obscuration due to the central beam hole on the mirrors. The key issue is how to eliminate the "forbidden pitch". It is impossible to completely avoid this problem only by projector design, while we can practically reduce impact on the projected pattern. To solve this problem, there will be three strategies:

(1) Make beam hole as small as possible.

(2) Optimize off-axis illumination.

(3) Optimize the partial coherence factor.

Figure 8 displays the central beam holes. In this paper, to distinguish between the obscuration factor and the partial coherence factor, we use the uppercase Greek letter $\Sigma$ as the obscuration factor and the lowercase letter $\sigma$ as the partial coherence factor.

The central holes are designed to pass the beam of NA 0.2, with a 2 mm gap surrounding the beam edge. Obscuration on AM1 is usually smaller than AM2, thus we discuss only AM2.

As shown in Fig. 8, the normalized hole sizes (obscuration factors) are

$$\Sigma_x = 0.13 \tag{15a}$$

$$\Sigma_y = 0.26 \tag{15b}$$

$\Sigma = 1$ represents the diffraction cone (the mirror diameter). NA is low thus the beam hole and the horizontal obscuration are small. We made the secondary mirror AM2 surface near the wafer to maintain Petzval-sum rule. This decision also helped to reduce the size of the beam hole.

We introduce quadrupole illumination, which can bypass the central obscuration. The logic pattern consists mainly of vertical and horizontal lines, whose diffractions are distributed along the horizontal and vertical axes, as shown in Figure 12). If the separation of quadruple illumination spots (0-th order diffractions) in horizontal and vertical is larger than the obscuration size, the diffractions will not be lost by the obsecrated hole. In the current design, it obviously satisfies the condition as follows.

$$W_{off\ axis} = 0.71 > \Sigma_x = 0.13 \tag{16a}$$

$$H_{off\ axis} = 0.71 > \Sigma_y = 0.26 \tag{16b}$$

In the special case of staggered contact holes [9], the diffraction pattern should have 60 degrees of rotational symmetry, and thus there remains a chance to hit the forbidden pitch in the obscured area. We can partially rescuer obscured spots by the partial coherent source. Because the partial coherence factor is larger than the obscuration, i.e. $\sigma_x = 0.25 > \Sigma_x = 0.13$, the diffraction spots will spread wider than the central hole width, and thus the dropped diffraction will be rescued, which is shown in Figs. 12 and 14..

Further investigation will be required, including source mask optimization (SMO) and optical proximity correction (OPC) using computational lithography.

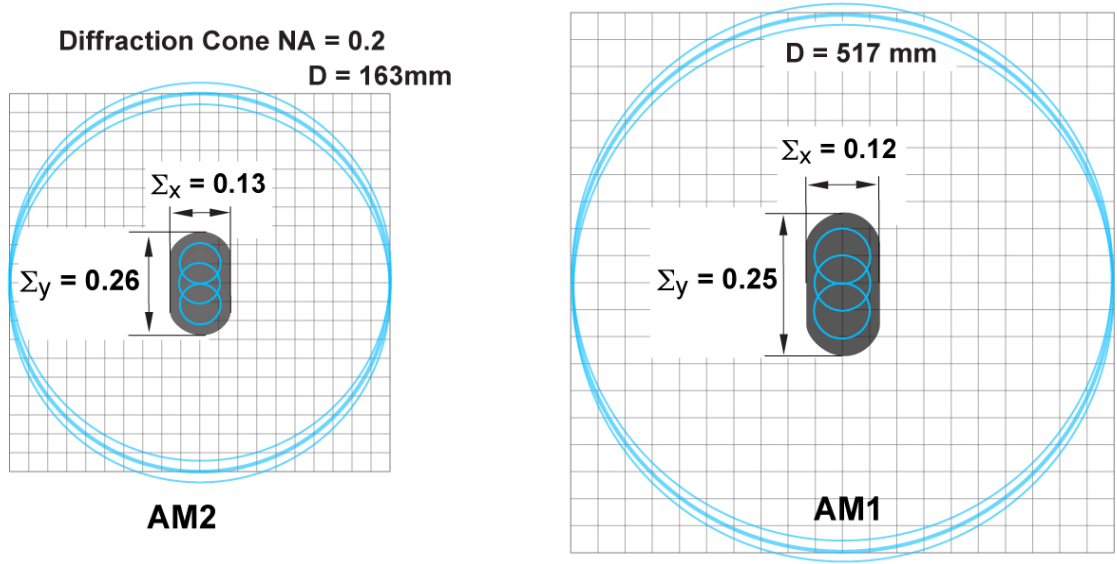

Figure 8. Beam obscuration on the primary and the secondary mirror. Beam holes are designed to match the beam edge of NA 0.2 with a 2 mm gap around the beam. Three circles indicate the diffraction cones on the axis and both field edges.

In order to circumvent the central obscuration and enhance spatial resolution, the EUV light is introduced in front of the mask through two narrow cylindrical mirrors positioned on either side of the diffraction cone. This provides average normal illumination and reduces the mask 3D effect. The simplified illumination system provides symmetrical quadrupole off-axis illumination, bypassing central obscuration and improving spatial resolution, and also realizes Köhler illumination. To avoid blocking diffraction by the cylindrical mirrors, a dual line field concept is introduced. The technical details are currently in the design phase and will be presented in a separate paper in the near future.

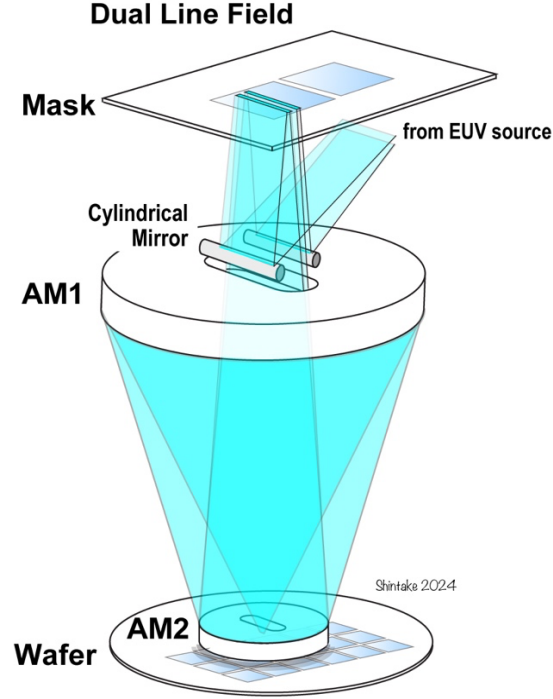

Figure 9. The illumination of the mask is provided by two cylindrical mirrors.

## 4.  PARTIAL COHERENT SOURCE

If we use point light source illumination, the frequency components close to the edge are sharply cut off by the aperture (cutting by hard edge), which often causes ringing tails on the image. To avoid this problem, a partial coherent source is commonly used in optical lithography.

The partial coherence factor is defined as follows[11].

$$\sigma = \frac{2 sin\theta_{max}}{2NA} = \frac{source\ diameter}{lens\ diameter} \tag{17}$$

In the case of a point light source, $\sigma = 0$. This is called Coherent Illumination.

In conventional UV lithography, the partial coherent illumination factor of 0.2 is commonly used in the case of quadrupole illumination, and it is desirable to use the same value for EUV. It is important to note that the partial illumination will also moderate impact of the central obscuration. As shown in Fig. 8, obscuration size in x-direction in $\Sigma_x = 0.13$, thus partial coherent light of $\sigma = 0.2$ will erases the hole.

We discuss the source size in x- and y-direction separately. First, x-direction is discussed. The scan field width is narrow: 2.5 mm wide, and thus the natural angular spread of the EUV plasma source satisfies the required angular spread. As shown in Fig. 10, we assume the collection angle of the segment mirror in the x-direction is 1 radian per mirror. The diameter of the tin plasma is about 100 um [13], from which we cut 50 um width, then magnified 50 times by the illuminator and delivered to the mask as a 2.5 mm wide line field. Since the phase space area is preserved through the linear optics (identical to the emittance conservation law in the particle accelerators), the angular divergence is adiabatically reduced by 1/50 and thus the angular spread on the mask becomes 20 mrad. Compared to the entrance pupil diameter of 2 x NA/m= 2 x 0.2/5 = 80 mrad, the partial coherent factor becomes $\sigma_x = 20/80 = 0.25$, which meets the required value.

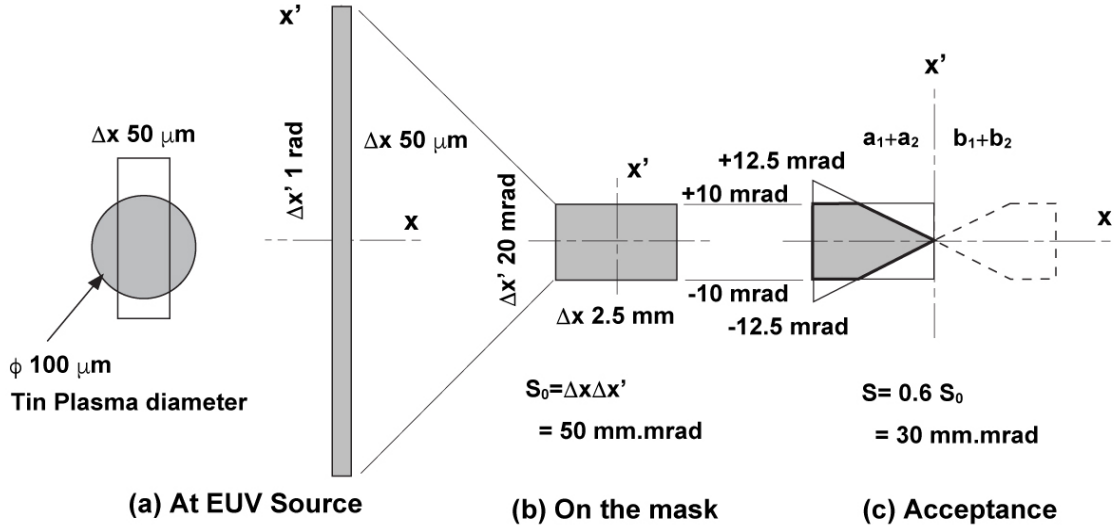

**(a) At EUV Source**     **(b) On the mask**     **(c) Acceptance**

Figure 10. The phase space distribution from the EUV source to the mask. (a) The diameter of the plasma is about 100 um, from which we cut 50 um width. (b) The source size is magnified 50 times through the illuminator and delivered to the mask as a 2.5 mm wide line field. The angular divergence is adiabatically reduced by 1/50. (c) The acceptance between two cylindrical mirros and the collimator cuts the phase space. 60 % of the photon flux can reach to the wafer. The dashed line shows the pair illumination.

In the y-direction, the illuminator expands the light into a wide line width of 100 mm to cover the size of the mask. However, this results in an extremely small angular divergence, which does not satisfy the required partial coherence. To increase the divergence in the y-direction, a 'ripple mirror' can be introduced in the illuminator. The 'ripple mirror' was originally introduced by Henry N. Chapman and Keith A. Nugent for a curved scan field in 1999. The mirror surface has periodic undulations that mix light rays, effectively increasing the partial coherence factor without substantial loss of light. Technical details are well described in the reference [12].

Figure 11 shows the off-axis quadrupole illumination pattern at the focal plane by taking account the partial coherence factors $\sigma_x = 0.25$, $\sigma_y = 0.2$. The shadow of two cylindrical mirrors is smeared and some of the illumination and diffraction can pass through, so there should be a 10~15% margin on the mirrors where the aberration should be corrected.

The four illumination spots are symmetrically distributed near pupil size at a 45-degree angle from the axis. The available resolution is determined by frequency width, which is $2NA\cos(45°) = 1.4NA$. Then, the critical dimension becomes,

$$CD = \frac{1}{2} \cdot \frac{\lambda}{2NA \cdot \cos(45°)} = k_1 \cdot \frac{\lambda}{NA} = 24 \text{ nm} \qquad (18a)$$

$$k_1 = \frac{1}{4\cos(45°)} = 0.35 \qquad (18b)$$

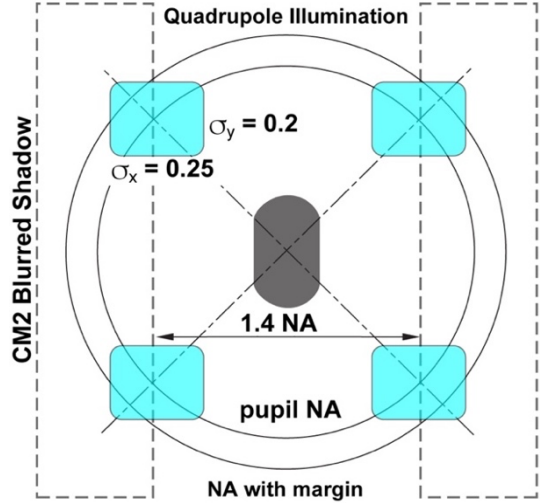

Figure 11. Off-axis quadrupole illumination pattern at the focal plane. By taking off-axis angle to meet the pupil size of NA, then the frequency span becomes 1.4 NA. The shadow of the cylindrical mirros is smeared and part of the illumination and the diffraction can pass through, thus 10~15% margin should be kept on the mirrors, i.e., aberration correction need to cover diffraction for NA 0.22.

## 5. IMAGING CAPABILITY WITH QUADRUPOLE ILLUMINATION

This section presents the preliminary results of the imaging analysis, which demonstrate the impact of central obscuration and shadow of two cylindrical mirrors under quadrupole illumination. Further optimization work will be required to investigate various logic patterns, as well as the spacing of two cylindrical mirrors and the partial coherent and pupil fill factors. This issue is related to the cost of photons lost at the line scan slit and central hole, which reflect back to the EUV source power requirement. The flexibility of the illumination scheme in this proposal is somewhat limited. However, we have sufficient EUV power at the projector to optimize (also pupil fill factor) for various logic patterns with reasonable contrast, which is sufficient for our purposes.

The Fresnel number $F$ is defined as $F = a^2/L\lambda$, where $a$ is the characteristic size, $L$ is the distance from the object and $\lambda$ is the incident wavelength. For 1 micron-meter field of the logic pattern, only 1 mm away from the mask surface, $F = 0.07 \ll 1$ and thus the diffraction becomes Fraunhofer regime, i.e., we can treat diffraction with Fourier transform. For the nanometer pattern, $F$ is always very small and in Fraunhofer regime.

As can be seen in Fig. 3, the diffraction propagates from AM2 to AM1, between which the light rays from different field heights cross at the focal plane, producing the Fourier pattern of the mask image. As shown in Figure 8, three circles indicate the diffraction cones on the axis and both field edges. The amount of displacement is relatively small compared to the diameter, so we approximate the imaging capability by diffraction from the field centre. The central obscuration is higher in AM2 than in AM1 as shown in Figure 8, so we estimate the obscuration at AM2.

Figure 12 shows the FFT analysis of the 27 nm HP vertical line. The images from left to right are the test pattern, diffraction at the focal plane, overlapped diffraction aligned 0th order to the origin and the back FFT aerial image. The second row is the case of no obscuration (no hole). A part of the diffraction is blocked with the cylindrical mirrors in front of the mask, while the shadows are not totally black. This is due to the non-zero source size, i.e., the phase space ($\Delta x$, $\Delta x'$) of $50 \ \mathrm{mm \cdot mrad}$ (see Fig. 10), which smears the shadow of the mirrors.

It is important to note that intense diffraction forms a rectangular shape, which is surrounding the central hole. Consequently, there is minimal power directed into the central hole. Therefore, the impact of the central obscuration is relatively insignificant in this case. This doesn't change for a horizontal line.

Figure 13 illustrates the intensity across the vertical line. It can be observed that the contrast is not diminished by the central obscuration; rather, it is slightly enhanced by it. This may be explained as follows. The loss of signal in the hole is equivalent to the addition of a signal with a 180-degree phase offset, which enhances contrast in a manner similar to that of a phase mask.

Figure 14 is FFT analysis of 35 nm stagged hole. The images from left to right are the test pattern, diffraction at the focal plane, overlapped diffraction aligned 0-th order to the origin and the back FFT aerial image. The second row is the case of no obscuration (no hole). Certain amount of the diffraction power directed into the hole, thus the contrast becomes lower as shown in Fig. 15. As discussed by Jo Finders et.al. [10], off-axis illumination with three or six poles is suitable for printing in higher resolution and better contrast. However, the system proposed in this paper is based on quadrupole illumination and may not be a best choice for printing staggered arrays. Despite this, the system is simple and economically feasible.

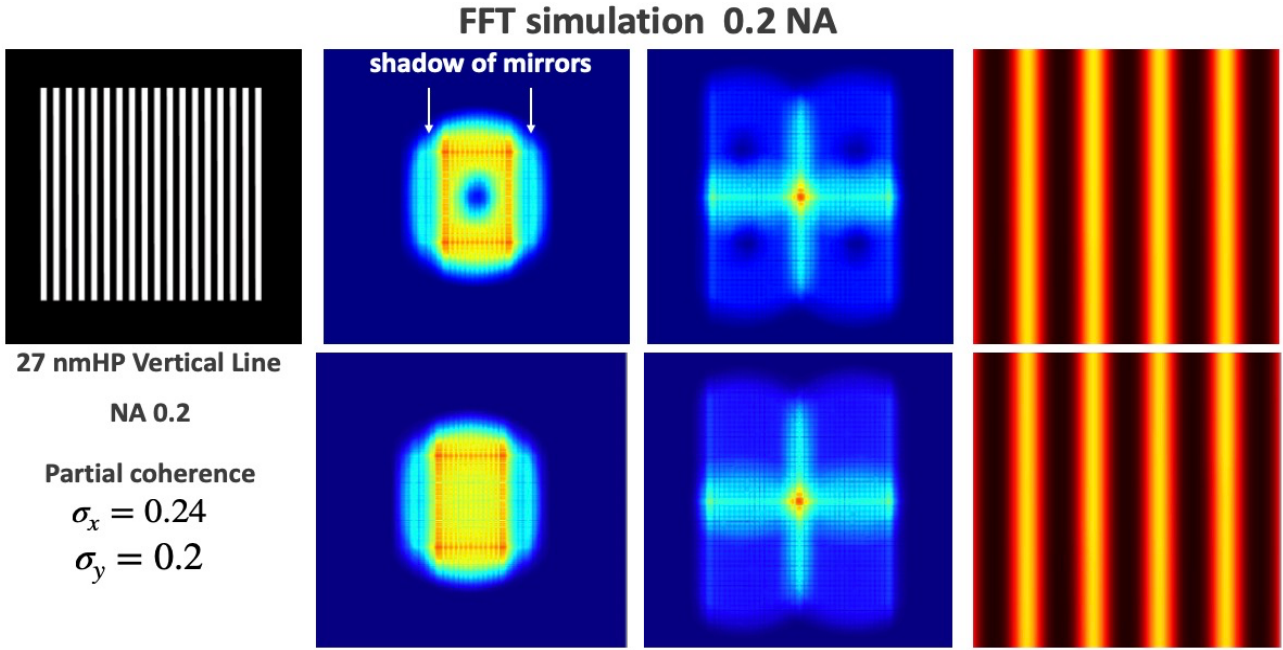

Figure 12. FFT analysis of the 27 nm HP vertical line. The images from left to right are the test pattern, diffraction at the focal plane, overlapped diffraction aligned 0-th order to the origin and the back FFT aerial image. The second row is the case of no obscuration (no hole).

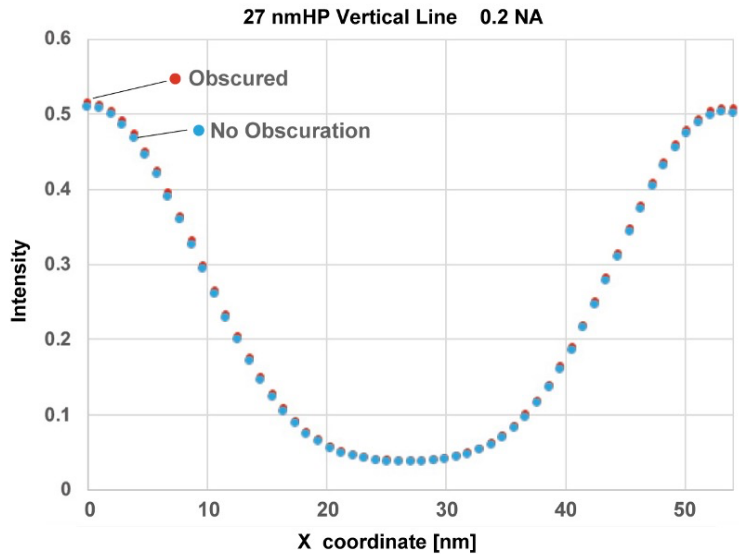

Figure 13. The intensity profile at the middle of the pattern.

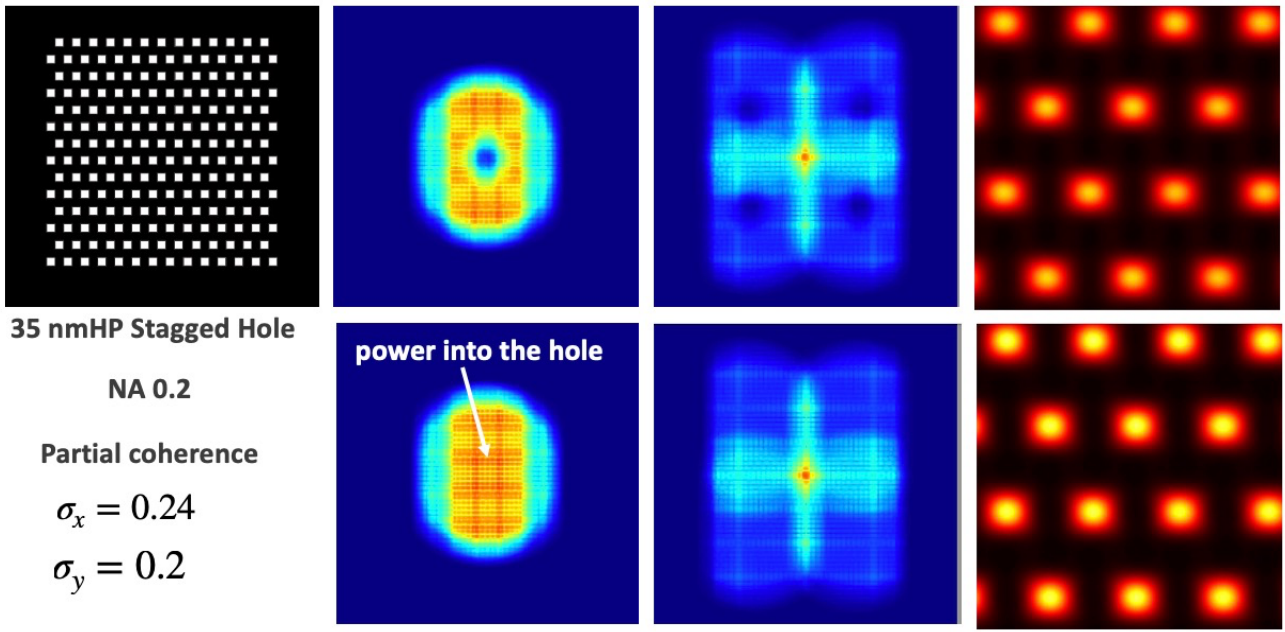

Figure 14. FFT analysis of 35 nm stagged hole. The images from left to right are the test pattern, diffraction at the focal plane, overlapped diffraction aligned 0-th order to the origin and the back FFT aerial image. The second row is the case of no obscuration (no hole).

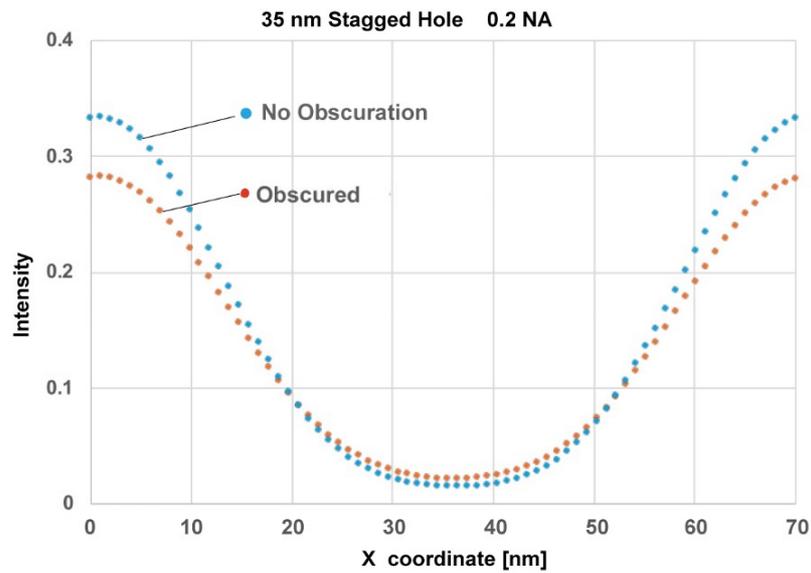

Figure 15. The intensity profile at the middle of the patter

## 6. SUMMARY AND COMMENTS

Key features of the proposed EUV lithography are

- Low power consumption < 1/10
  (AC power consumption of 1 MW in current system will be reduced to 100 kW)

- EUV source hardware will be simpler, resulting in lower cost and longer lifetime.

- Simple two-mirror projector lowers the capital cost, makes design reliable.

- Easier maintenance.

To go narrow line.

- Multiple patterning with 24 nm HP (0.2NA, 20 mm field) can be a reasonable strategy. It should be noted that lithography costs will be lower.

- 0.3 NA (16 nm HP, 10 mm field) can be realized with curved surface mask.
  10 mm x 26 mm field size is suitable for mobile applications.

- Combination with "chiplet" design is perfect match.

 

The author proposes that a proof-of-principle experiment should be conducted as soon as possible, potentially in a half-scale model, namely OID 1000 mm, 0.2 NA, 10 mm field with and without a curved surface mask.